\documentclass[12pt]{iopart}
\usepackage{amssymb}
\usepackage{graphicx}
\usepackage{dcolumn}
\usepackage{bm}
\usepackage{epsfig}
\usepackage{color}
\usepackage{iopams}
\usepackage{cases}
\newcommand{\beq}{\begin{equation}}
\newcommand{\eeq}{\end{equation}}
\newcommand{\beqa}{\begin{eqnarray}}
\newcommand{\eeqa}{\end{eqnarray}}
\newcommand{\ea}{\end{array}}

\begin{document}

\title{Effective-range signatures in quasi-1D matter waves:
sound velocity and solitons}

\author{F. Sgarlata$^{1}$,  G. Mazzarella$^{1,2}$, L. Salasnich$^{1,2,3}$}

\address{$^{1}$Dipartimento di Fisica e Astronomia``Galileo Galilei'',
Universit\`a di Padova, Via F. Marzolo 8 - 35131 Padova, Italy \\
$^{2}$Consorzio Interuniversitario per le Scienze Fisiche
della Materia (CNISM)\\
$^{3}$Istituto Nazionale di Ottica (INO) del Consiglio Nazionale
delle Ricerche (CNR), Sezione di Sesto Fiorentino, Via Nello Carrara,
1 - 50019 Sesto Fiorentino, Italy}

\date{\today}

\begin{abstract}
We investigate ultracold and dilute bosonic atoms under
strong transverse harmonic confinement by using
a 1D modified Gross-Pitaevskii equation (1D MGPE), which accounts
for the energy dependence of the two-body scattering amplitude
within an effective-range expansion.
We study sound waves and solitons of the quasi-1D system
comparing 1D MGPE results with the 1D GPE ones.
We point out that, when the finite-size nature of the interaction is
taken into account, the speed of sound and the density profiles of
both dark and bright solitons show relevant quantitative
changes with respect to what predicted by the standard 1D GPE.
\end{abstract}


\maketitle

\section{Introduction}

The Gross-Pitaevskii equation (GPE), which plays a relevant role in the
study of Bose-Einstein condensates (BECs) made
of ultracold and dilute alkali-metal atoms, is based on the assumption
of a zero-range inter-atomic potential \cite{leggettreview2001}.
Recently, several experiments \cite{feshbe}
employing the Fano-Feshbach resonance technique in cold atomic
collisions \cite{tie92} have shown
that it is possible to change the magnitude and the sign of the
scattering length $a_s$ by using an external magnetic field.
Thus, by using Fano-Feshbach resonances
it is now possible to explore, at fixed density $n$, regimes
where the GPE and its assumptions lose their validity.

In this work, going beyond the Fermi pseudopotential
approximation (contact interaction) of the standard GPE,
we focus on sound waves and solitons in a BEC of
interacting bosons at zero temperature under a strong transverse harmonic
confinement. We take into account the dependence on the energy
of the two-body scattering amplitude employing
the effective-range expansion illustrated by Fu and {\it et al.}
in \cite{gaoPRA2003} by inserting therein the correction proposed by
Collin and co-workers in \cite{pethickPRA2007}.
These two ingredients allow us to write a modified version of the
Gross-Pitaevskii equation (MGPE, as named in (\cite{gaoPRA2003})) which
incorporates the finite-range nature of the inter-atomic interaction.
We reduce the dimensionality of the 3D MGPE by integrating out
the degrees of freedom in the radial plane and we obtain a 1D
MGPE which takes into account both the scattering length
and the effective range of the inter-atomic potential.
We model the boson-boson interaction by means of three potentials:
hard-sphere potential, Van-der-Waals potential, and square-well potential.
We set the s-wave scattering length to a given value and calculate,
for this $a_s$, the effective range of each above model potential.
In this way, we find relevant quantitative changes
of the atomic cloud properties, i.e. the speed
of sound and the width of the dark and bright solitons, with respect
to the results provided by the familiar one-dimensional
Gross-Pitaevskii equation.

\section{The modified Gross-Pitaevskii equation}

We consider $N$ interacting bosons of mass $m$ confined by an external
trapping potential $V_{trap}(\vec{r})$ at zero temperature.
The Hamiltonian is then given by
\beq
\label{globalhamiltonian}
H = \sum_{i=1}^N h(\vec{r}_i) +\frac{1}{2}\sum_{i=1}^N\sum_{j \neq i}
V(\vec{r}_i-\vec{r}_j)
\;,\eeq
where
\beq
\label{singleparticlehamiltonian}
h(\vec{r}_i) = -\frac{\hbar^2}{2m}\nabla_i^2 + V_{trap}(\vec{r}_i)
\eeq
with $V(\vec{r}_i-\vec{r}_j)$ describing the interaction between
two bosons at positions $\vec{r}_i$ and $\vec{r}_j$.
The ground-state properties of a weakly interacting bosonic gas can be
very efficiently described by using the standard Gross-Pitaevskii
equation (GPE) \cite{leggettreview2001}. As well known, one can
derive the GPE minimizing the GP
energy functional $E_{GP}$. Describing the inter-atomic potential
by the Fermi pseudopotential
\beq
\label{pseudo}
V_F(\vec{r}_i-\vec{r}_j)=g\delta(\vec{r}_i-\vec{r}_j)
\;,\eeq
where the coupling strength $g$ is
\beq
\label{g}
g=\frac{4\pi \hbar^2 a_s}{m}
\;,\eeq
with $a_s$ the interparticle s-wave scattering length,
the energy functional $E_{GP}$ reads:
\beq
\label{functionaldelta}
E_{GP}[\phi,\phi^*]=N \int d^3\vec{r}\ \phi(\vec{r})^*h(\vec{r})
\phi(\vec{r})+\frac{g}{2}N(N-1)\int d^3\vec{r} |\phi(r)|^4
\;,\eeq
where $\phi(\vec{r})$ is the single-particle wave function
(all the $N$ bosons are in the same single-particle state).
By exploiting the variational approach, where the functional $E_{GP}$
is required to have a minimum with respect to $\phi(\vec{r})$ obeying
the normalization condition:
\beq
\label{normalization}
\int d^3\vec{r}\,|\phi(\vec{r})|^2 = 1
\;,\eeq
by using that for very large $N$ one can write that $(N-1)\sim N$ and
by employing the Lagrange multipliers method, one arrives to the standard GPE
\beq
\label{gpe}
\bigg[-\frac{\hbar^2}{2m} \nabla^2+V_{trap}(\vec{r})+g\,N|
\phi(\vec{r})|^2\bigg]\,\phi(\vec{r})= \mu \, \phi(\vec{r}) \; ,
\eeq
where $\mu$ is the chemical potential.

At this point some considerations about the inter-atomic
potential (\ref{pseudo}) are in order. Such a potential
ignores completely the dependence on
the energy of the scattering amplitude. This approximation, however,
is valid provided $na_{s}^{3}$ is sufficiently small.
On the other hand, for stronger confinements
and larger values of $na_{s}^{3}$, a better treatment of atomic interactions
that preserves much of the structure of the GP theory is possible.
This goal can be pursued by introducing an effective interaction
potential $V_{eff}$ which gives the energy dependence of the
scattering amplitude through an effective-range expansion
which will also depend on the effective range $r_e$ of the inter-atomic
potential \cite{gaoPRA2003,pethickPRA2007}. Specifically, in the following,
we use the effective interaction potential
\beq
\label{effpotential}
V_{eff}(\vec{r}_i-\vec{r}_j)=V_F(\vec{r}_i-\vec{r}_j)
+V_{mod}(\vec{r}_i-\vec{r}_j) \; ,
\eeq
where
\beq
\label{vmod}
V_{mod}(\vec{r}_i-\vec{r}_j)=\frac{g_2}{2}[\delta(\vec{r}_i-\vec{r}_j)
\nabla^2_{\vec{r}_i-\vec{r}_j}+\nabla^2_{\vec{r}_i-\vec{r}_j}
\delta(\vec{r}_i-\vec{r}_j)]
\eeq
and
\beq
\label{g2}
g_2={4\pi\hbar^2\over m} a_s^2 \left( \frac{1}{3}a_s-\frac{1}{2}r_e \right)
\; .
\eeq
In this case, from Eq. (\ref{effpotential}), it can be
deduced that the energy functional has an extra term $E_{mod}$,
due to $V_{mod}$, having the following form:
\beqa
&&E_{mod}[\phi^*,\phi] \simeq \frac{N}{2}\int d^3\vec{r}_1\int d^3
\vec{r}_2 \phi^{*}(\vec{r}_1)\phi^{*}(\vec{r}_2)V_{mod}(\vec{r}_1-\vec{r}_2)
\phi(\vec{r}_1)\phi(\vec{r}_2)=\nonumber\\
&=& \frac{N}{2}\int d^3\vec{R}  \int d^3\vec{r} \phi^{*}(\vec{R}
+\frac{\vec{r}}{2})\phi^{*}(\vec{R}-\frac{\vec{r}}{2}) V_{mod}(\vec{r})
\phi(\vec{R}+\frac{\vec{r}}{2})\phi(\vec{R}-\frac{\vec{r}}{2})
\label{vmedioHmod}
\;,\eeqa
where we have made use of $(N-1) \sim N$ and the second row is
a re-writing of the first one in the two body center-of-mass frame
($\vec{r}=\vec{r}_i-\vec{r}_j$, $\vec{R}=(\vec{r}_i+\vec{r}_j)/2$).
The simplification of $E_{mod}$ achieved by doing calculations in the above
frame and minimization of the (inclusive-$E_{mod}$) modified
Gross-Pitaevskii (MGP) energy functional
\beq
\label{functionalplusmod}
E_{MGP}[\phi^*,\phi] = \int d^3\vec{r} \phi^*\left[ -\frac{\hbar^2}
{2m}\nabla^2 + V_{trap}(\vec{r})+\frac{g}{2}\left| \phi \right|^2
+ \frac{g_2}{4}\nabla^2\left(\left|\phi\right|^2\right)\right]\phi
\;\eeq
with respect to $\phi^*$ with the constraint (\ref{normalization})
provide the following modified Gross-Pitaevskii equation (MGPE)
\beq
\label{mgpe}
\bigg[-\frac{\hbar^2}{2m} \nabla^2+V_{trap}(\vec{r})+g\,N|\phi(\vec{r})|^2
+\frac{N}{2}g_2\nabla^2(|\phi(\vec{r})|^2)\bigg]\,\phi(\vec{r})=
\mu\,\phi(\vec{r})\; .
\eeq
Notice that a similar nonlinear Schr\"odinger equation has been derived and
studied by Garc\'{\i}a-Ripoll, Konotop, Malomed,
and P\'{e}rez-Garc\'{\i}a \cite{boris-rompi}. Their investigation starts
from the Hartee equation for bosons, which is a nonlocal integral
Schr\"odinger equation (nonlocal GPE) \cite{sala-old},
and it is based on a gradient expansion of the nonlocal
GPE \cite{boris-rompi,sala-old}.

\section{The one-dimensional MGPE}

We assume that the external confinement potential $V_{trap}(\vec{r})$
is obtained by superimposing to a very strong isotropic harmonic confinement
in the $x-y$ (radial) plane a generic shallow potential along the $z$
(axial) direction, so that
\beq
\label{exttrapping}
V_{trap}(\vec{r}) = \frac{1}{2}m\omega_\perp^2(x^2+y^2)+U(z)
\;,\eeq
where $\omega_\perp$ is the trapping harmonic frequency.
The spatial degree of freedom in the radial plane is thus frozen and
the system can be considered, in practice, one-dimensional (1D)
in the axial direction. As suggested by
the form (\ref{exttrapping}) of the external trapping potential,
we shall use  the following Gaussian ansatz for the single-particle
wave function $\phi(\vec{r})$:
\beq
\label{gaussianansatz}
\phi(\vec{r}) = \frac{\varphi(z)}{\sqrt{\pi}a_\perp} \,
e^{-\frac{x^2+y^2}{2a^2_\perp}} \;,
\eeq
where $a_\perp = \sqrt{\hbar/(m\omega_\perp)}$ is the transverse
characteristic length of the ground state of the harmonic potential and
$\int dz |\varphi(z)|^2=1$. This ansatz will be valid when
$g|\varphi|^2/2\pi a_{\bot}^2\ll 2 \hbar \omega_{\perp}$ \cite{salasnichPRA2002}.
Inserting Eqs. (\ref{exttrapping}) and (\ref{gaussianansatz}) into
Eq. (\ref{functionalplusmod}) and then minimizing with respect to
$\varphi^*$ leads to the 1D version of the modified Gross-Pitaveskii equation
\beq
\label{mgpe1d}
\left[ -\frac{\hbar^2}{2m}\frac{d^2}{dz^2}+U(z) +
\gamma\left|\varphi\right|^2+\frac{1}{2}\gamma_2\frac{d^2}{dz^2}
\left| \varphi\right|^2\right] \varphi(z) =\tilde{\mu}\varphi(z)
\;,\eeq
where
\beq
\label{gammainteraction}
\gamma = \frac{1}{2\pi a_\perp^2}\left(g - {g_2\over a_\perp^2} \right)
\; , \qquad
\gamma_2 = \frac{g_2}{2\pi a_\perp^2}
\; , \qquad
\tilde{\mu} = \mu-\hbar\omega_\perp \; .
\eeq
The effective-range effects heralded by Eq. (\ref{mgpe1d}) become
clear when the ratio of the absolute value of the
effective range $|r_e|$ to the inter-atomic distance
(referred to the 3D system) is of the same order of magnitude of
the ratio of this distance to the absolute value of the s-wave scattering
length $|a_s|$. In this situation the dependence on the energy of
the two-body scattering amplitude (see, for example,
\cite{gaoPRA2003,scatteringbook}) cannot be neglected and the usual
1D GPE is not able to describe adequately anymore the physics of our system.
Thus, to study the effects of the finite-size nature of the boson-boson
interaction on the atomic cloud properties, in our forthcoming 1D
MGPE-based studies, $|r_e|$ and $|a_s|$ will be chosen in such a way
to meet the condition mentioned above. Moreover, note that results
from Eq. (\ref{mgpe1d}) are reliable as long as $N|a_s|/a_\perp \ll 1$.

\section{Interaction potentials}

In this section we present three toy models for the two-body interaction
potential between atoms. Then, we shall use these three potentials
in the analysis of the sound velocity and solitonic
waves within the system under investigation.

\begin{itemize}
\item
{\it Hard-sphere potential}. This model for the description of
the boson-boson interaction is defined as follows
\beq
V(r) =
\infty\quad r\leq a_s,\quad while \quad
V(r)=0 \quad r>a_s
\label{hs}
\;.\eeq
For this potential,
\beq
\label{rehs}
r_e = \frac{2}{3} a_s
\;,\eeq
and one thus reduces to the standard GPE since $\gamma _2=0$,
as it can be seen from see the first and second formula of
Eq. (\ref{gammainteraction}) with $g_2$ given by Eq. (\ref{g2}).

\item {\it Square-well potential}. In this case, the two-body collisions
are described by a potential well characterized by a finite depth:
\beq
V(r) =
-V_0\quad r\leq r_0, \quad while \quad
V(r)=0  \quad r> r_0
\label{step}
\eeq
with $V_0$ positive. It is possible to show that in the limit of
sufficiently small incident wave vector ($q \rightarrow 0$),
the s-wave scattering length $a_s$ is given by
\beq
\label{aswell}
a_s= r_0\left[ 1-\frac{\tan\left(\chi(0) r_0\right) }{\chi(0) r_0}\right]
\;,\eeq
and the effective range $r_e$ by
\beq
\label{rewell}
r_e = r_0\left[ 1-\frac{r_0^2}{3a_s^2}-\frac{1}{\chi(0)^2 a_s r_0} \right]
\;,\eeq
where $\chi(0)^2=mV_0/\hbar^2$.

\item {\it Van-der-Waals potential}. When the interaction is Van
der Waals-like, the interaction potential may be approximated by
a potential well for $r <r_0$ (this latter being called empty-core radius),
while by a function of the form $-C_6/r^6$ otherwise, that is
\beq
V(r) =
\infty \quad r\leq r_0, \quad while \quad
V(r)=-C_6/r^6 \quad r> r_0
\label{vdW}
\;,\eeq
where $C_6$ is a parameter which quantifies the interaction strength.
Note that the potential above is reminiscent of the Ashcroft
pseudopotential used to treat conduction electrons in alkali metals.
For the potential (\ref{vdW}) the s-wave scattering length $a_s$ and
the effective range $r_e$ have the following expressions \cite{gaoPRA1998}:
\beq
\label{asvdw}
a_s = \frac{\Gamma^2\left(\frac{3}{4}\right)}{\pi}
\left( 1-\tan\Phi\right) l_{vd}
\;,\eeq
\beq
\label{revdw}
r_e = \frac{2\pi}{3\Gamma^2\left(\frac{3}{4}\right)}
\frac{1+\tan^{2}\Phi}{(1-\tan\Phi)^2}l_{vd}
\;,\eeq
respectively. In the above formulas $l_{vd}$ is a $C_6$-dependent
characteristic length and $\Phi$ a function depending on
the ratio $l_{vd}^{2}/r_0$:
\beqa
\label{lvdphi}
l_{vd} = \left(\frac{mC_6}{\hbar^2}\right)^{1/4} \qquad \Phi
= \frac{l^2_{vd}}{2r_0^2} -\frac{3\pi}{8} \;.
\eeqa

\end{itemize}

The forthcoming analysis will be focused on the sound velocity and
solitonic density profiles for each of the three boson-boson interaction
potential models above presented. We keep fixed the scattering length
$a_s$ and calculate the effective interaction range $r_e$ by using
the formulas above provided,  that is, Eq. (\ref{rehs}) for the hard
spheres potential (\ref{hs}), Eqs. (\ref{aswell}) and (\ref{rewell})
for a given $V_0$ in the case of the square-well potential (\ref{step}), and
Eqs. (\ref{asvdw}) and (\ref{revdw}) for a given $C_6$ in the case
of the Van-der-Waals potential (\ref{vdW}).

\section{Sound velocity}

We want to gain physical insight both in the spatial and temporal evolution
of our system. The theoretical tool which permits us to do this
is the time-dependent version of the modified one-dimensional
Gross-Pitaevskii equation (\ref{mgpe1d}). We suppose that $U(z)=0$, and scale lengths, times, and energies in units of $a_{\perp}$,
$1/\omega_{\perp}$, and $\hbar \omega_{\perp}$, respectively. We use thus the following adimensional time-dependent 1D MGPE:
\beq
\label{timedp1dmgpe}
i\frac{\partial}{\partial t } \varphi(z,t) =
\left[-\frac{1}{2}\frac{d^2}{dz^2}+\gamma\left|\varphi\right|^2
+ \frac{1}{2}\gamma_2\frac{d^2}{dz^2}\left| \varphi\right|^2\right]
\varphi(z,t) \;,
\eeq
where, for simplicity of notation, we have denoted the dimensionless quantities by the same symbols used for those with dimensions.
We are interested, in particular, in the consequences of a perturbation,
with respect to the equilibrium, created at a given spatial point of
the system at a given time. We start writing $\varphi(z,t)$ as:
\beq
\label{polar}
\varphi(z,t) = \sqrt{n(z,t)}e^{iS(z,t)}
\;,\eeq
with $n(z,t)$ describing the density profile and $S(z,t)$ related
to the velocity field $v(z,t)$ via the relation
\beq
\label{velocity}
v(z,t) =\frac{\partial}{\partial z} S(z,t)
\;.\eeq
By inserting the two equations above in the time-dependent 1D MGPE
(\ref{timedp1dmgpe}), one obtains the hydrodynamic equations (HEs)
\beqa
\label{hydrodynamicequations}
&&\frac{\partial v}{\partial t} + \frac{d}{dz}\left[ \frac{1}{2}v^2 +
\gamma n + \left(\gamma_2-\frac{1}{4n}\right)\frac{d^2}{dz^2}n +
\frac{1}{8n}\left( \frac{dn}{dz} \right)^2 \right] = 0 \nonumber\\
&&\frac{\partial n}{\partial t} + \frac{d}{dz}(nv) = 0
\;.\eeqa

At this point, let us suppose to perturb the system with respect to the
equilibrium configuration characterized by $n(z,t)=n_{0}$ and $v(z,t)=v_{0}=0$:
\beqa
\label{perturbation}
&&n(z,t)=n_{0}+ \delta n(z,t)\nonumber\\
&&v(z,t)=v_{0}+\delta v(z,t)
\;.\eeqa
We use these formulas
in the hydrodynamic equations (\ref{hydrodynamicequations})
and assume to be in the stationary regime, $v_0=0$. Under the hypothesis
that the perturbation is sufficiently weak so as to retain only the
$\delta n$-first-order terms in the HEs, we get
\beq
\label{deltanequation}
\frac{\partial^2}{\partial t^2}\delta n -n_0\,\gamma\frac{d^2}{dz^2}(\delta n)
-n_0\left(\gamma_2-\frac{1}{4n_0}\right)\frac{d^4}{dz^4}\delta n = 0
\;.\eeq
If the perturbation is a plane wave, that is $\delta n(z,t)=
Ae^{i(k_z z-\omega t)}+A^{*}e^{-i(k_z z-\omega t)}$, the relation of dispersion
which characterizes the oscillations associated to the wave induced by
the perturbation is
\beq
\label{dispersionrelation}
\omega = k\sqrt{n_0\, \gamma - \left( n_0\,\gamma_2-\frac{1}{4}\right)k^2}
\eeq
which depends on the equilibrium density $n_0$ and contains information
about two-body collisions via $\gamma$ and $\gamma_2$, see the first
two formulas of Eq. (\ref{gammainteraction}), and Eqs. (\ref{g}) and
(\ref{g2}). The perturbation will stable with respect to time for real
$\omega$ that is always guaranteed when $a_s=2/3 r_e$. If this is the case,
the dispersion relation (\ref{dispersionrelation}) is the usual Bogoliubov
dispersion, that is
\beq
\omega^2=\frac{k^2}{2}\left(\frac{k^2}{2}+2c_s^2
\right)
\label{realomega}
\eeq
which, in the limit of sufficiently small wave vector ($k\rightarrow 0$)
gives back the usual dispersion relation of the sound wave, that is
\beq
\omega=c_s k
\;\eeq
with the velocity $c_s=\sqrt{n_0\gamma}$ of sound propagating in the system related to the interaction parameters, equilibrium density, and harmonic trap characteristics. To see more clearly such a dependence we use the standard units of measure so that one has
\beq
\label{soundvelocity}
c_{s}^2= n_0\,{2\hbar^2 a_s\over m^{2} a_{\bot}^2}\left(1 - {1\over 3}
{a_s^2\over a_{\bot}^2} + {1\over 2} {r_e \, a_s\over a_{\bot}^2} \right) \; ,
\eeq
where we have take into account the definitions of $\gamma$, $g$ and $g_2$.

As above commented, we study the sound velocity $c_s$ as a function of
the equilibrium density $n_0$, Eq. (\ref{soundvelocity}), and analyze
such a quantity for each of three interaction potentials previously described.

\begin{figure}[htpb]
\centering
\includegraphics[scale=0.9]{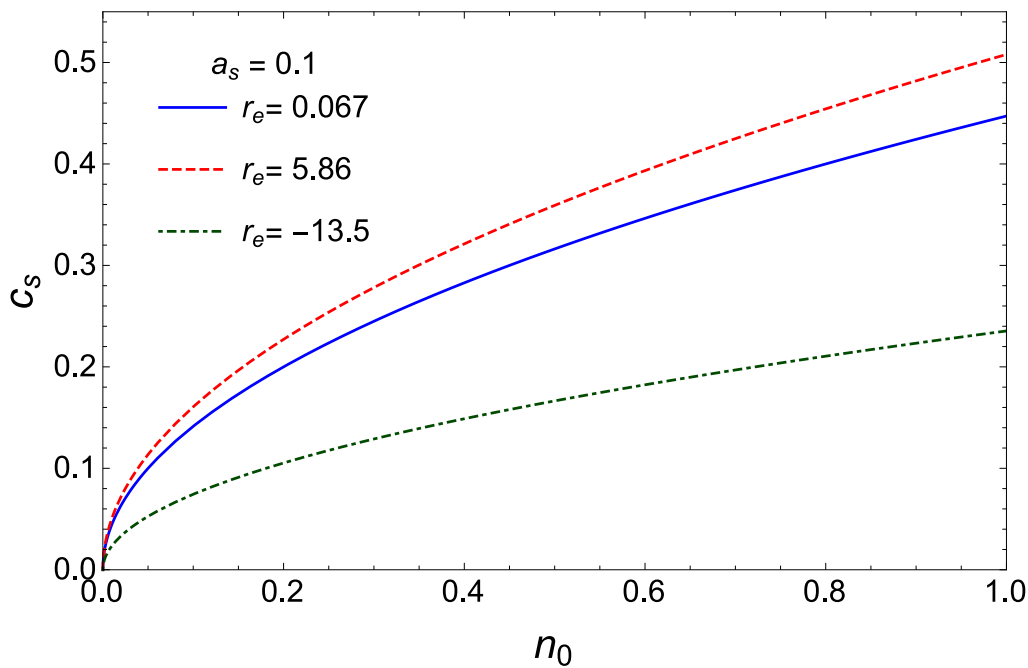}
\caption{Sound velocity $c_s$ vs axial equilibrium density $n_0$
for $a_s=0.1$. Solid line: hard-sphere potential (\ref{hs})
[this curve is the same provided
by the standard 1D GPE]. Dot-dashed line: square-well potential
(\ref{step}) [$r_0 = 0.8$, $V_0 = 31.05$].
Dashed line: Van-der-Waals potential (\ref{vdW})
[$C_6 = 0.07$, $r_0 = 0.278$]. Lengths in units of $a_{\perp}$, times
in units of $1/\omega_{\perp}$, $c_s$ in units of $a_\perp \omega_\perp$,
$n_0$ in units of $1/a_{\perp}$, $C_6$ in units of
$\hbar \omega_\perp a_{\perp}^6$.}
\label{fig:confrontosuono}
\end{figure}

Fig. 1 shows the sound velocity $c_s$ as a function of the axial
equilibrium density $n_0$ on varying the shape of the inter-atomic
interaction potential, see Sec. 4. We have fixed the s-wave scattering
length $a_s$ and calculated [given $r_0$ and $V_0$ for the potential
(\ref{step}) and $C_6$ and $r_0$ for the potential (\ref{vdW})]
the value of $r_e$ for each inter-atomic potential by using Eq. (\ref{rehs})
for the hard-sphere potential, Eqs.(\ref{aswell})-(\ref{rewell})
for the square-well potential, and Eqs.(\ref{asvdw})-(\ref{revdw}) for the
Van-der-Waals potential.

For any chosen set of parameters of the inter-atomic
potential under investigation the final result
will only depend on the obtained value of $a_s$ and $r_e$. Clearly,
except the case of the hard-core potential,
fixing $a_s$ several parameters of the inter-atomic potential under
investigation will give the same $r_e$ and the same sound velocity $c_s$.

We observe that the behavior of the sound
velocity, when the type of boson-boson interaction changes, is qualitatively
the same. However, at a given $n_0$, by increasing $\gamma_2>0$
one gets a larger sound velocity $c_s$.

The solid line of Fig.1 represents the
sound velocity as a function of the axial equilibrium density when
the interaction between the bosonic atoms is described by the hard-sphere
potential (\ref{rehs}). Since $r_e=2/3 a_s$ - Eq. (\ref{rehs})
- $\gamma_2=0$ (see Eq. (\ref{g2}) and the third formula of
Eq. (\ref{gammainteraction})) so that one reduces to the same behavior
predicted by the 1D GPE with a Dirac-delta interaction characterized
by the assigned $a_s$, see Eq. (\ref{timedp1dmgpe}).

For instance, Fig. 1 compares sound velocity versus density
in the three potentials of interest. We can thus conclude
that the finite-size nature of the inter-atomic interaction has
the effect to produce quantitative changes in the behavior of the sound
velocity $c_s$ with respect to that predicted by the familiar 1D GPE.

\section{Solitons}

We start by considering the time-dependent 1D MGPE (\ref{timedp1dmgpe}).
When $\gamma_2=0$ we reduce to the standard time-dependent one-dimensional
Gross-Pitaevskii equation. It is well known that this equation admits
the possibility of studying topological configurations of the
Bose-Einstein condensate like solitonic solutions
(solitary waves preserving their form and propagating
with a constant velocity $v$) with positive (repulsive inter-atomic
interaction) or negative (attractive inter-atomic interaction)
s-wave scattering length $a_s$ \cite{soliton-books}
\beq
\label{solitonwaves}
\varphi(z,t) =f(z-vt)e^{iv(z-vt)}e^{i\left(\frac{1}{2}v^2-\mu\right)t}
\;.\eeq
The solutions corresponding to $a_s>0$ are the dark solitons.
The axial density $|f|^2$ of these solitons assumes the same finite value
when $x \rightarrow \pm \infty$ (with $x=z-vt$ the comoving coordinate of
the soliton) and is characterized by an hole-structure with a minimum
at $x=0$. The difference between the phases of the wave function at
$\pm \infty$ is finite. For $a_s<0$ one has the bright solitons that
set up when the negative inter-atomic energy of the BEC balances the
positive kinetic energy so that the BEC is self-trapped in the axial
direction. In this case $|f|^2$ goes to zero when $x \rightarrow \pm \infty$
and exhibits a pulse-structure with a maximum at $x=0$.
The difference between the phases of the wave function
at $\pm \infty$ is zero.

We focus on solitary waves when the
the effective-range correction is taken into account, that is
with $\gamma_2$ finite. Proceeding thus from the 1D MGPE, we look
for its solutions of the form (\ref{solitonwaves}) which inserted
in Eq. (\ref{timedp1dmgpe}) provide the following differential equation:
\beq
\label{de}
-\frac{1}{2} f''+ \gamma f^3 + \frac{1}{2}\gamma_2 \left(f^2\right)''f = \mu f
\;,\eeq
where $'' \equiv \displaystyle{\frac{\partial^2}{\partial x^2}}$.
We observe (see the discussion in the sequel) that 1D MGPE admits dark
(bright) solitonic solutions when the nonlinearity $\gamma$ is positive
(negative). Therefore, due to the form of $\gamma$ - first formula of
Eq. (\ref{gammainteraction}) - it is possible to have a given type of
soliton irrespective of the sign of $a_s$.

\subsection{Dark Solitons}

We study the black solitons that are dark solitons characterized by a
vanishing axial density at $x=0$ and zero velocity $v$ with respect to
the condensate.  It is possible to achieve a relation which implicitly
defines the solution $f$ of the differential equation (\ref{de}) that reads
\beq
\label{implicitdark}
\sqrt{1-2\gamma_2f(z)^2} arctanh(\frac{f(z)}{f_\infty}) =
\sqrt{\gamma}f_\infty z
\eeq
with $f_\infty$ being the absolute value got by $f$ at $\pm \infty$
and $\gamma>0$. Since $0<|f(z)|^2<1$, the dark solitons solution exists
when $-\infty < \gamma_2<1/2$.

The density profile $f(z)^2$ can be thus studied as a function of the
axial coordinate $z$ by solving numerically Eq. (\ref{implicitdark})
when one knows the features of the boson-boson interaction, i.e.
both $\gamma$ and $\gamma_2$. To set these two quantities, we have
followed the same procedure followed to obtain Fig.1 (see Sec. 5).
We have thus plotted $f(z)^2$ versus $z$, Fig.2.

We observe that when one takes into account the finite-size nature of
the inter-atomic interaction, the width of the solitary wave under
investigation is qualitatively the same of that one would found by using
the familiar one-dimensional Gross-Pitaevskii equation (solid line,
see the discussion in Sec. 5) but its magnitude meaningfully changes
with respect to the latter case.

\begin{figure}[htpb]
\centering
\includegraphics[scale=0.9]{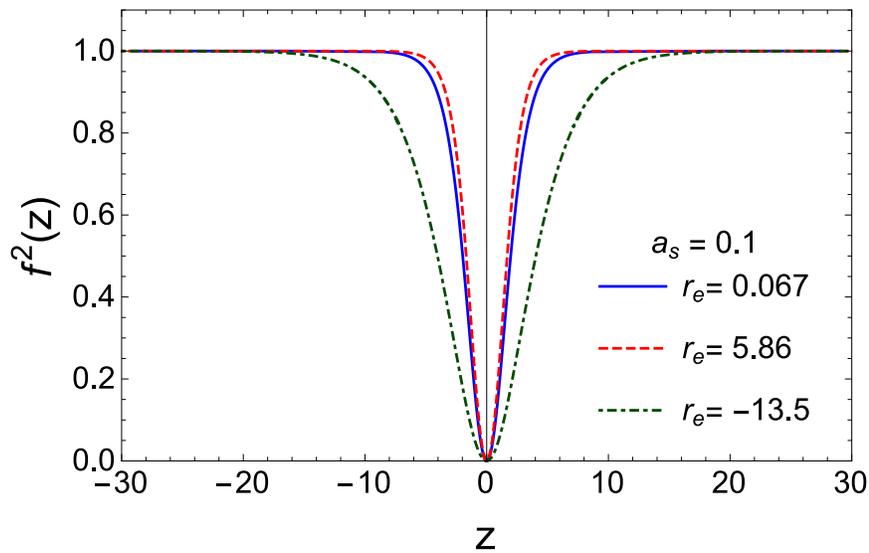}
\caption{Axial density profile $f(z)^2$ of the black soliton vs axial
coordinate $z$ for $a_s=0.1$. Solid line: hard-sphere potential (\ref{hs})
[this curve is the same provided by the standard 1D GPE].
Dot-dashed line: square-well potential
(\ref{step}) [$r_0 = 0.8$, $V_0 = 31.05$].
Dashed line: Van-der-Waals potential (\ref{vdW})
[$C_6 = 0.07$, $r_0 = 0.278$]. Lengths in units of $a_{\perp}$, energies
in units of $\hbar \omega_\perp$, $C_6$ in units of
$\hbar \omega_\perp a_{\perp}^6$, $f(z)^2$ in arbitrary units.}
\label{fig:blacksoliton}
\end{figure}

Actually, the width $\Delta z$ at
half-minimum of the dark soliton can be easily calculated from
Eq. (\ref{implicitdark}) setting $f_{\infty}=1$, $f(z)=1/2$, and
$z=\Delta z/2$. In this way we immediately find
\beq
\Delta z = {2\over arctanh({1\over 2})}
\sqrt{1-{1\over 2} \gamma_2\over \gamma} \; .
\label{senzadime}
\eeq
Taking into account the definitions of $\gamma$ and $\gamma_2$,
Eq. (\ref{gammainteraction}) with Eqs. (\ref{g}) and (\ref{g2}), this
formula gives the width $\Delta z$ of dark solitons
as a function of the scattering length $a_s$, effective range $r_e$, and
transverse width $a_{\bot}$ of the harmonic confinement.

\begin{figure}[htpb]
\centering
\includegraphics[scale=0.9]{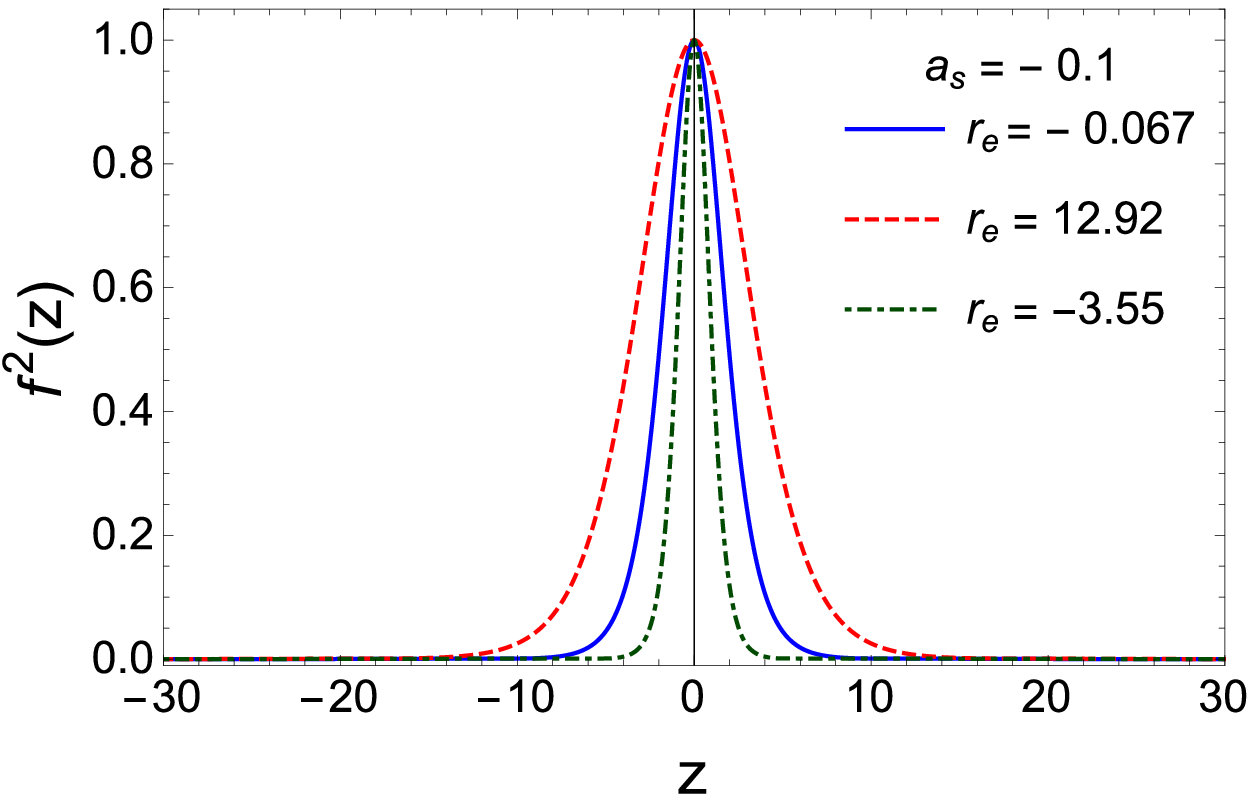}
\caption{Axial density profile $f(z)^2$ (at $t=0$) of the bright soliton
vs axial coordinate $z$ for $a_s=-0.1$. Solid line: hard-sphere
potential (\ref{hs}) [this curve
is the same provided by the standard 1D GPE].
Dot-dashed line: square-well potential (\ref{step})
[$r_0 = 0.5$, $V_0 = 82.1011$]. Dashed line: Van-der-Waals potential
(\ref{vdW}) [$C_6 = 0.07$, $r_0 = 0.2492$]. Lengths in units of
$a_{\perp}$, energies in units of $\hbar \omega_\perp$, $C_6$ in
units of $\hbar \omega_\perp a_{\perp}^6$, $f(z)^2$ in arbitrary units.}
\label{fig:brightsoliton}
\end{figure}

\subsection{Bright Solitons}

We start from Eq. (\ref{de}). When $\gamma<0$, the constant of motion
for this equation is
\beq
\label{constant}
K = \frac{1}{2}(f')^2 + \mu f^2 -\frac{1}{2}\gamma f^4 - \frac{1}{4}
\gamma_2 \left[ \left( f^2 \right)' \right]^2
\;.\eeq
By requiring that $f$ and its first derivative tend to
zero at $\pm \infty$, we get $K=0$. By imposing that $f$ is
maximum for $x=0$, we obtain $\mu = -\frac{1}{2}|\gamma|f(0)^2$,
and by defining $f=\phi(x)^{1/2}$  we get, from Eq. (\ref{constant}),
\beq
\phi'=\pm \sqrt{\frac{8(K-\mu\phi+\frac{1}{2}\gamma\phi^2)}
{\left( \frac{1}{\phi} - 2\gamma_2 \right)}}
\;.\eeq
Then, by integrating the above expression with $+$ and by using
$K=0$ and $\mu=-1/2 \gamma |f(0)|^2$, one has that
\beq
\label{intebright}
2\sqrt{|\gamma|}z = \int_{f(z)^2}^{f(0)^2} dy \sqrt{\frac{1-2\gamma_2y}
{y^2(f(0)^2-y)}} \; .
\eeq

\begin{figure}[htpb]
\centering
\includegraphics[scale=0.5]{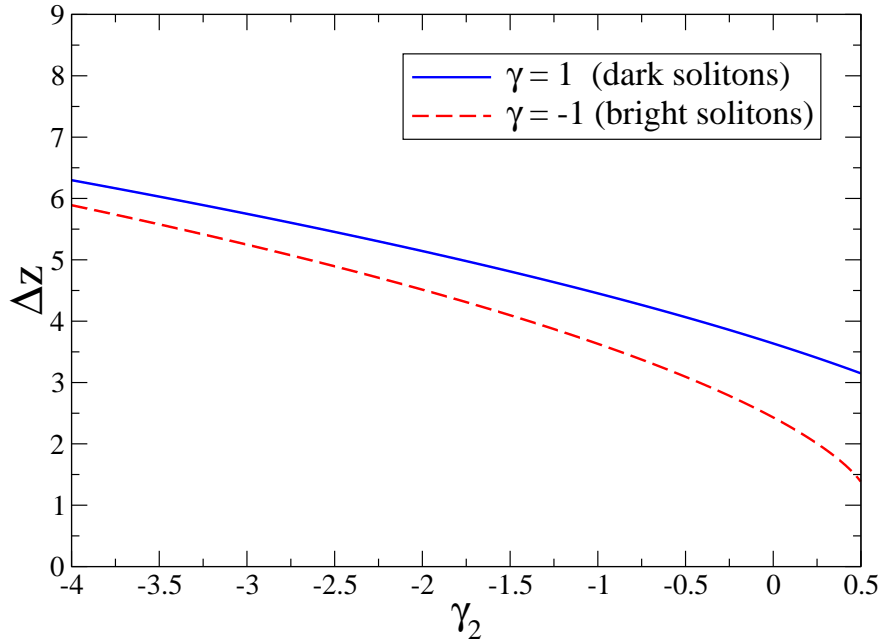}
\caption{Width $\Delta z$ of dark solitons (solid line) and bright solitons
(dashed line) as a function of the coupling $\gamma_2$. We set $\gamma=1$
for dark solitons and $\gamma=-1$ for bright solitons. $\Delta z$ in units of $a_{\bot}$, $\gamma$ in units of $\hbar \omega_{\bot} a_{\bot}$, 
$\gamma_2$ in units of $\hbar \omega_{\bot} a_{\bot}^3$.}
\label{fig:widths}
\end{figure}

The integral at the right-hand side of Eq. (\ref{intebright})
can be numerically solved by allowing for a study of the density
profile $f(z)^2$ of the soliton as a function of the axial coordinate
$z$ setting both $\gamma$ and $\gamma_2$. Therefore for the bright
solitons as well, we have studied the density profile $f(z)^2$ as
a function of the axial coordinate varying the boson-boson interaction
potential by following the same path as for the black solitons.
These results are enclosed in Fig. 3. From the plots therein,
it can be observed - as for the sound velocity and the dark solitons -
that the width is quantitatively affected by the nature of the inter-atomic
interaction potential. The width $\Delta z$ at
half-maximum of the bright soliton can be calculated from
Eq. (\ref{intebright}) setting $f(0)=1$, $f(z)=1/2$, and
$z=\Delta z/2$. In this way we immediately find
\beq
\Delta z = {1\over \sqrt{|\gamma|}}
\int_{1/4}^{1} dy \sqrt{\frac{1-2\gamma_2 y}{y^2(1-y)}} \; .
\label{senzadime2}
\eeq
This formula is more complex than Eq. (\ref{senzadime}), but Fig.
\ref{fig:widths} shows that Eq. (\ref{senzadime}) has the same
behavior of Eq. (\ref{senzadime2}) once the signs
of $\gamma$ are taken into account.

\section{Conclusions}

We have considered a system of interacting atomic bosons confined in
a strong harmonic confinement in the radial plane plus a weak potential
along the axial direction at zero temperature. We have carried out our
analysis going beyond the Fermi pseudopotential approximation and described
the gas evolution by employing a modified one-dimensional Gross-Pitaevskii
equation (1D MGPE) in the absence of the axial potential. By using the
latter equation we have studied the propagation of sound waves and that
of solitons in the system under investigation. We have used the 1D MGPE
to study the sound velocity versus the axial density and the density
profiles of the solitons (black and bright) as function of the axial
coordinate by modeling the boson-boson interaction via an hard-sphere
potential, a square-well potential, and a Van-der-Waals potential.
We have performed our investigations by fixing the s-wave scattering
length $a_s$ and calculating the effective-range $r_e$ corresponding,
for this $a_s$, to each inter-atomic potential. This analysis has
allowed us to conclude that the effective-range signatures reflect
in important quantitative changes (with respect to the results of
the familiar 1D GPE) of the speed of sound and solitary waves density profile.

GM and LS acknowledge financial support from MIUR
(PRIN Grant no. 2010LLKJBX).

\end{document}